# Effect of YIG Nanoparticle Size and Clustering in Proximity-Induced Magnetism in Graphene/YIG Composite Probed with Magnetoimpedance Sensors: Towards Improved Functionality, Sensitivity and Proximity Detection

S. Hosseinzadeh[1], L. Jamilpanah[2], J. Shoa e Gharehbagh[2], M. Behboudnia[1], S. M. Mohseni[2,*]

[1] Department of Physics, Urmia University of Technology, Urmia, Iran
[2] Faculty of Physics, Shahid Beheshti University, Evin, 19839 Tehran, Iran

*Proximity-induced magnetism (PIM) in graphene (Gr) adjacent to magnetic specimen has raised great fundamental interests. The subject is under debate and yet no application is proposed and granted. In this paper, toward accomplishment of fundamental facts, we first explore the effect of particle size and clustering in the PIM in Gr nanoplates (GNPs)/yttrium iron garnet (YIG) magnetic nanoparticle (MNP) composite. Microscopic analyzes suggest that fine MNPs distributed uniformly on the GNPs have higher saturation magnetization due to the PIM in Gr. We propose that such magnetic plates can thus be used to shield the stray field generated on the surface of magnetic sensors and play a role as a magnetic lens to prevent the field emanating outside the body of magnetic specimen. The GNPs/YIG composites are coated on a magnetic ribbon and proposed for application in magneto-impedance (MI) sensors. We show that such planar magnetic flakes enhance the MI response against the external applied magnetic field significantly. The suggested application can be furthermore developed toward bio-sensing and magnetic shielding in different magnetic sensors and devices.*

***Keyword:*** *proximity induced magnetism; magnetic graphene; magnetic sensor; magneto-impedance*

[*]Corresponding author's email address: m-mohseni@sbu.ac.ir, majidmohseni@gmail.com

Tel: +989354880368

## 1. Introduction

Proximity-induced magnetism (PIM) is a process where a non-magnetic material acquires magnetization due to coupling with a magnetic film[1]. The first report on magnetic proximity was broadcasted in 1969[2] in superconductors. It has been found that the superconducting transition temperature in Pb/Pd/Fe structure decreases with decreasing the thickness of Pd, indicating that the Pd layer became magnetized in contact with the Fe layer. Later, it has been shown that Fe and Co ferromagnetic materials can induce magnetization into 4d and 5d elements such as Pd and Pt.[3-6] Very recently, the scenario revived and wider windows of materials have represented counterintuitively PIM. The Graphene (Gr) layer transferred on an electrically insulator yttrium iron garnet (YIG) thin film illustrates magnetic signal in the Hall Effect (AHE). The non-magnetic Gr layer has become magnetized while sitting on magnetic YIG thin film in a YIG/Gr bilayer.[7] This subject opened a field in view of deep understanding of the PIM in two-dimensional materials (2DM) family with many questions and proposals.[8-15]

There are several studies suggesting that the interface between the ferromagnetic and non-magnetic materials is the key toward observing PIM.[16-18] Different approaches to achieve room temperature ferromagnetism in Gr have been reported by using magnetic nanoparticles (MNPs) with focus on the surface shape of matrix and the uniformity of MNPs on the surface of Gr multilayer.[19] However, study on particle size and distribution on the surface of Gr plates is rare. In this paper, we report evidence of induced ferromagnetism at room temperature in graphene nanoplates (GNPs) with layer number less than three that is decorated by YIG-MNPs.

Technical application of magnetic Gr that is magnetized via the PIM is rare. Such magnetic plate can have alternative benefits compared to known MNPs, as they are being frequently used. The planar shape of Gr with high surface to volume ratio, while being magnetized, can be applied as magnetic shielding to absorb undesired magnetic field presented in different devices such as sensors. Nonetheless, such MNPs on the surface of Gr plates that provides PIM, occupy a little portion of the Gr plates surface. Hence, together with other well-known Gr functionalities, such PIM in Gr can be applied in many systems, such as bio-sensors and many other magnetic functional elements.

Here, we introduce technological application of such magnetic GNPs. Magnetic properties of such planar magnetic plates is employed to improve the sensing of the magnetoimpedance (MI) effect through their magnetic shielding ability. High sensitivity and facile technological requirements have presented the MI effect a rich research field.[20, 21] This effect is the change in electrical impedance against external DC magnetic field. The MI is determined through the skin depth ($\delta$), $\delta = (\rho/\pi\mu_t f)^{1/2}$, of the high frequency ($f$) current and the transvers magnetic permeability ($\mu_t$) of metallic ferromagnet with electric resistivity ($\rho$).[22] Impedance of a metallic ferromagnet changes by the new skin depth of the current when external magnetic field is applied. Fundamental prospective of ferromagnetic metals and development of highly sensitive magnetic field sensors has increased interest in MI effect.[23-25] Therefore, the impedance of the ribbon is a function of frequency of driving current and external dc magnetic field ($H$) through $\mu_t$ and $\delta$. At high frequencies, the skin depth $\delta$ decreases and so the current passes at the sheath of the ribbon and so the electrical and magnetic environmental conditions would highly affect the MI behavior.[26] This phenomenon has two prospects, one is related to the magnetic field sensor performance and another one is related to the environmental functionality response. There are reports on the magnetic field sensitivity enhancement by coating layers with different magnetization and conductivities on the surface of MI sensors.[27-32] They can tune the MI response mainly due to closure of magnetic flux path at the surface of the MI elements. Interestingly, MI sensing element preserves as a surface media to probe the spin-orbit torque due to non-magnetic Pt[33] and IrMn[34] layer, as the thin skin depth is quite sensitive against tiny changes at the surface. Recently, we presented surface modification of MI sensor made of magnetic ribbons for environmental sensitivity and stability by coating vertical-Gr-oxide (GO)[32]. Essentially, Gr based materials play an important role in environmental sensitivity with preserving stability in different environments.[35, 36] Since the PIM influences the magnetization of the GNPs, these wholly magnetized plates can be mounted on the surface at the close proximity to the MI sensors. The MI sensor is not only affected by the magnetization of these plates, but it can be influenced by their shielding performance thanks to their planar shapes. We show that the MI response increases significantly at the proximity of the magnetic-GNPs composites coated on surface of the sensor.

## 2. Experimental

### 2.1. Materials

GNPs (N002-PDR, XY=7μm, z=50-100nm) is supplied by Angstron materials Inc. Ferric nitrate (Fe(NO$_3$)$_3$·9H$_2$O), yttrium nitrate (Y(NO$_3$)$_3$·6H$_2$O), citric acid, ethylene glycol, dimethylformamide (DMF) were all from Merck (99.9% pure) to prepare YIG MNPs.

### 2.2. Preparation of YIG MNPs

Following our previous work on the preparation of MNPs [37-40], YIG-MNPs were synthesized by citrate-nitrate (CN) and modified co-precipitation (MCP), as two different sets. For the CN synthesis, we dissolved the required amount of the metal nitrates in stoichiometric ratio of Y: Fe = 3:5 in distilled water. Citric acid was then added into the prepared aqueous solution to pH=1. The solution was heated to dry and then annealed in ambient air at the temperature of 800 °C with a heating rate of rate of 10 °C/min for 2hrs. For synthesis of YIG-MNPs by MCP, we mixed the required amount of Y and Fe nitrates in stoichiometric ratio of Y: Fe = 3:5 in DMF to form metal-organic solution. Ethylene glycol was then added into prepared metal-organic solution. A small amount of ammonia was added to the solution to adjust pH value to about 10.5. During the procedure, the precipitate continuously stirred using a magnetic agitator. Then the precipitate collected and washed with distilled water and ethanol. The collected precipitate dissolved in distilled water with small amount of citric acid to reach pH=2. Finally, the solution precursor heated to dry and annealed to 700 °C for 2 h. Details of the MCP method is well-discussed in our recent work.[41]

### 2.3. Preparing Gr/YIG:

0.015 g YIG powder from each set was added to 35 ml ethanol followed by probe sonication for 30 min. 0.015 g of GNPs was added to the mixture followed by additional sonication for 30 min. After the ultrasonic treatment, the solution was heated inside an oven at 75 °C for 12 h to dry the sample.

### 2.4. MI setup

Conventional melt-spinning technique was used to fabricate Co$_{68.15}$Fe$_{4.35}$Si$_{12.5}$B$_{15}$ magnetic ribbons with 0.8 mm width, 40 mm length and about 28 μm thickness. The impedance measurement was done using four-point probe method. An AC current passed through longitudinal direction of the ribbon (length= 4 cm) with different frequencies supplied by function generator with constant amplitude of 30 mA. The impedance was evaluated by measuring the voltage and current across the sample using a digital oscilloscope. An external magnetic field was applied along the ribbon axis to perform MI measurements. This magnetic field was produced by passing electrical current in a 40 cm long solenoid, which can generate a magnetic field up to 120 Oe. The longitudinal direction of samples was set perpendicular to the Earth's magnetic field to minimize its undesired impact on the measurements.

## 3. Results and discussion

### 3.1. Characteristics of YIG

**Figure 1** shows transmission electron microscopy (TEM) images of YIG-MNPs prepared by CN and MCP methods. It can be seen in **Figure 1**a that the size distribution of YIG-MNPs prepared by CN method is from 20-50 nm and it reveals that particles are aggregated and exhibit irregular shapes without shaped borders. Because of the observed aggregation in TEM images of this sample, the dynamic light scattering (DLS) results show a much larger dimension for these particles. **Figure 1**b illustrates TEM images of YIG-MNPs synthesized by MCP method, demonstrating a moderate clustering of particles. We can observe some small aggregations, which are composed of primary particles with the size distribution of 10-20 nm, matches the result of DLS measurement. **Figure 1**b describes in accordance with the calculated values for crystal size by the Scherrer's equation,[42] the mean diameter of YIG-MNPs prepared by MCP is 17 nm, which agrees with the values calculated from x-ray diffraction (XRD), that we recently reported in ref [41].

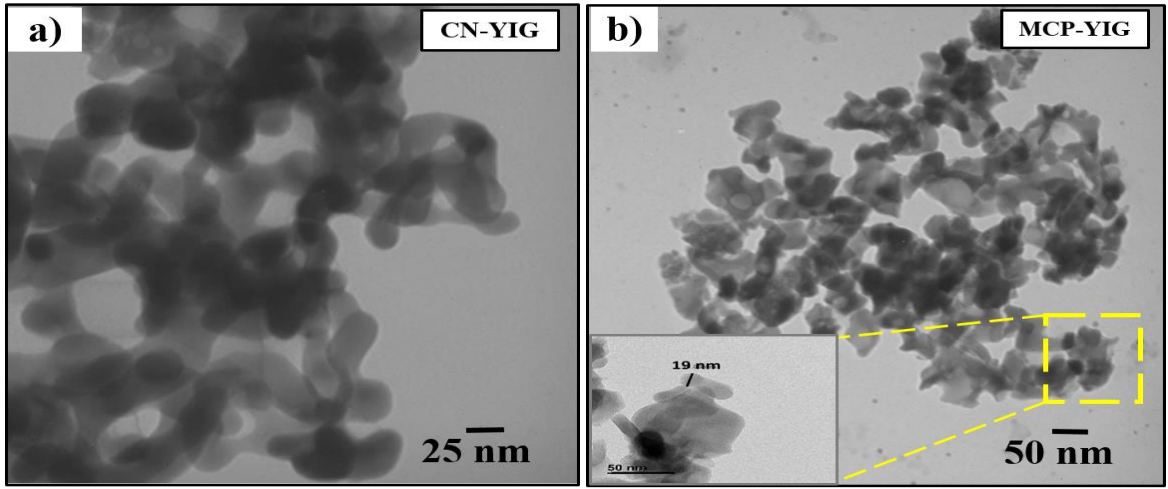

Figure 1. a) TEM image of YIG-MNPs prepared by CN that particles are aggregated and exhibit irregular shapes without shaped borders and the size distribution of YIG-MNPs is 20-50 nm, b) TEM and size distribution from DLS of YIG-MNPs synthesized by MCP method showing moderate clustering of particles and some small aggregations are observable.

**Figure 2**a shows XRD pattern of YIG for CN and MCP prepared samples at the room temperature. Samples Prepared by these methods completely contain YIG phase and no trace of intermediated phases is found. The mean crystallite sizes of YIG synthesized using CN at 800 °C and MCP at 700 °C were estimated to be about 38 and 17 nm, respectively. The details of phases have been described in our previous work**. [41] Figure 2**b shows the vibrating-sample magnetometry (VSM) recorded at room temperature for both samples. The saturation magnetization ($M_S$) of CN and MCP prepared samples is seen to be ~23.23 emu/g. The coercivity ($H_c$) and remanent magnetization ($M_r$) of CN sample are 30.23 Oe and 9.94 emu/g and those of MCP prepared samples are 30.1 Oe and 4.52 emu/g, respectively.

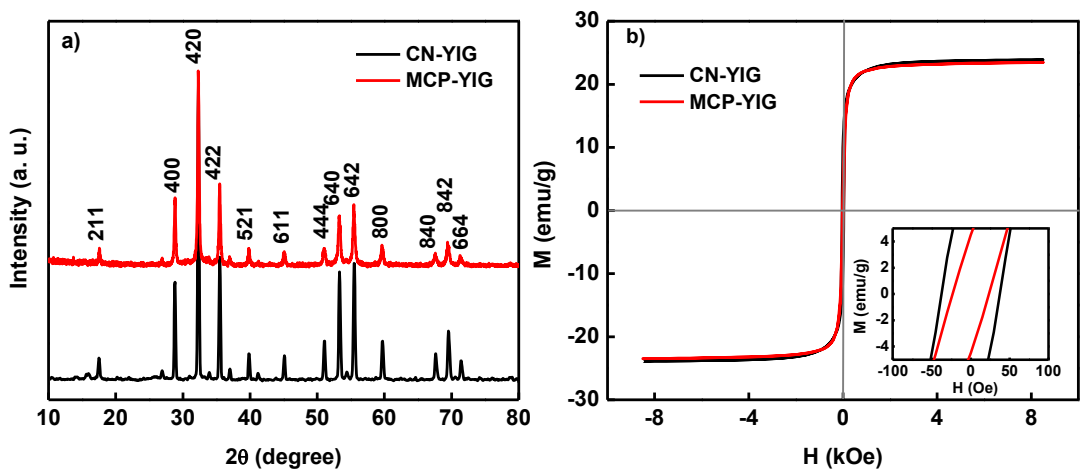

Figure 2. a) the XRD pattern of the YIG samples prepared via CN and MCP methods and b) the VSM hysteresis loops of them (inset shows the coercivity of the samples).

## 3.2. Characteristics of GNPs/YIG composite

**Figure 3**a, b shows the TEM and HR-TEM of GNPs decorated with two different YIG-MNPs products. As can be seen, the flake-like shapes of GNPs are clearly observed to be embedded with uniformly YIG-MNPs. A well distribution of YIG-MNPs can be seen in contact to GNPs in the sample prepared by MCP method and there is significant portion of the YIG attached on the surface of GNPs contrary to that made by CN method. Such a fine YIG-MNPs can enhance interface contact and facilitates the uniform attachment of MNPs on the surface and thus placed in close proximity to GNPs surface.

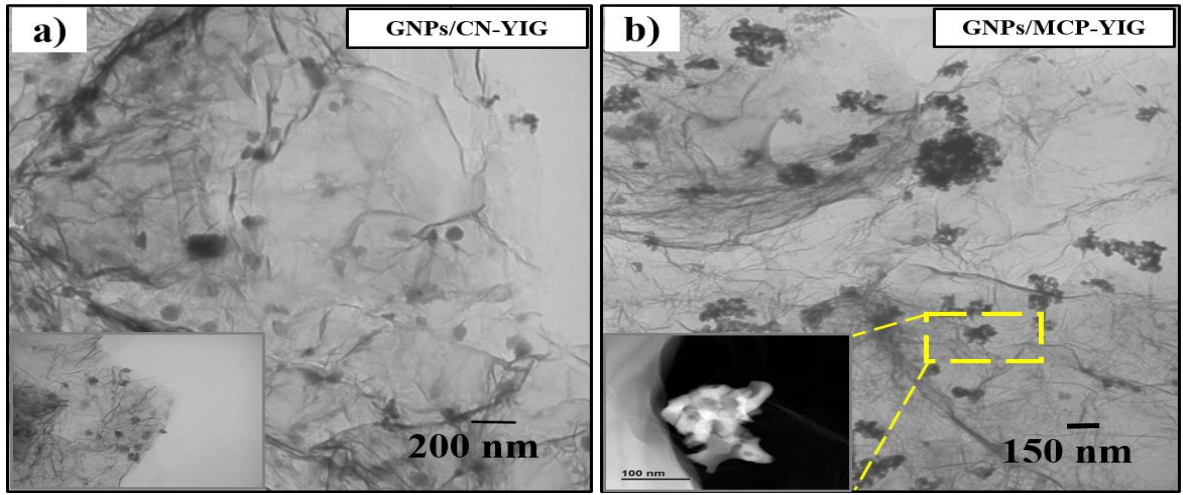

Figure 3. TEM images of the GNPs/YIG-MNPs composite with YIG prepared via a) CN and b) MCP methods. In both cases the flake-like shapes of GNPs are clearly observed to be embedded with uniformly YIG-MNPs.

X-ray diffraction patterns of GNPs decorated by MCP YIG-MNPs are presented in **Figure 4**a. The planes (002) at $2\theta= 25°$ attributed to GNPs and the main peaks of YIG-MNPs are observed too. In addition, GNPs/YIG profile indicates the formation of composites having main diffraction YIG peaks of (400), (420), (422), (44), (640) and (642) which can be indexed to JCPDS card no. 43-0507 properly dispersed in GNPs matrix. The sharp peaks of YIG MNPs confirm that their good crystallinity has not been destroyed during synthesis process. The other less noticeable peak of GNPs around $2\theta= 42°$ interrelated to the (100)/(101) plane is hinder by the stronger (521) peak of YIG and thus not observe properly.

**Figure 4**b shows the hysteresis loops recorded at room temperature for both samples. The $M_S$ of GNPs decorated by CN and MCP prepared samples by the mass ratio of 1:0.5 for the GNPs/YIG MNPs are 2.25 and 7.26 emu/g, respectively. The magnetic properties of hybrid materials can be tuned by making changes in proportion of MNPs to GNPs owing to the fact that magnetization may decrease by addition in GNPs composite, as non-magnetic portion.[30] Since ratio of GNPs to YIG is the same for both samples, the difference in saturation magnetization is due to contribution of size and uniformity of YIG MNPs. As we can see, the $M_S$ of GNPs in proximity to smaller YIG-MNPs is significantly higher than those prepared by larger MNPs. Thus, the induced ferromagnetism is observed in GNPs coupled to magnetic YIG-NPs via PIM of smaller MNPs.

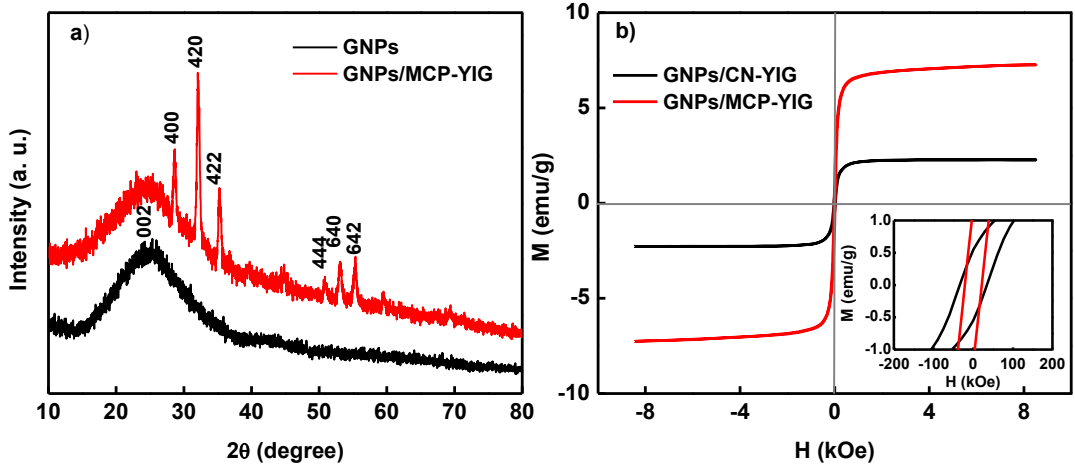

Figure 4. a) the XRD pattern of the GNPs/YIG composites prepared via MCP method and b) the VSM hysteresis loops of GNPs/YIG samples prepared via CN and MCP methods (inset shows the coercivity of the samples).

### 3.3. MI sensing of Gr/YIG composite

In order to understand the impact of GNPs/YIG on the MI response, MI ratio of the samples were measured at different frequencies of 2.5, 5, 7.5, 10, 12.5 and 15 MHz. The magnetic field was applied up to 120 Oe during MI measurements. The MI ratio can be defined as

$$MI\% = \frac{Z(H)-Z(H_{max})}{Z(H_{max})} \times 100 \quad (2)$$

where $Z$ refers to the impedance as a function of external field ($H$) and $H_{max}$ is the maximum field applied to the samples in the MI measurement. MI response for the bare ribbon and the ribbons drop coated by GNPs/CN-YIG and GNPs/MCP-YIG are presented in **Figure 5**. Field dependent MI ratio of the bare ribbon at different frequencies can be seen in panel (a) of **Figure 5**. There is a peak at low external applied magnetic fields because of transverse alignment of magnetic anisotropy of the ribbon against applied magnetic field direction.[43] In addition, this peak appears at the ribbon samples drop coated by GNPs/YIG composites. This means that the coating did not change the transverse anisotropy of the ribbon.

The maximum MI ratio occurs at the frequency of 10 MHz Relative contributions of domain wall motion and magnetization rotation to the transverse permeability should be considered in interpreting this trend.[22] The MI ratio increases by increasing frequency up to 10 MHz and then decreases by further increasing the frequency. The reduction of MI ratio at high frequencies is due to presence of eddy currents that causes damping of domain wall displacements and only rotation of magnetic moments takes place. In turn, the transverse magnetic permeability diminishes, and the MI ratio decreases.[22, 44] Maximum of the MI ratio for bare ribbon and ribbon coated by GNPs/YIG composites at the frequency range of 2.5-15 MHz are presented in **Figure 5**d. We observed that the increase of maximum MI ratio for the ribbons deposited by GNP/MCP-YIG is more than that of GNPs/NC-YIG.

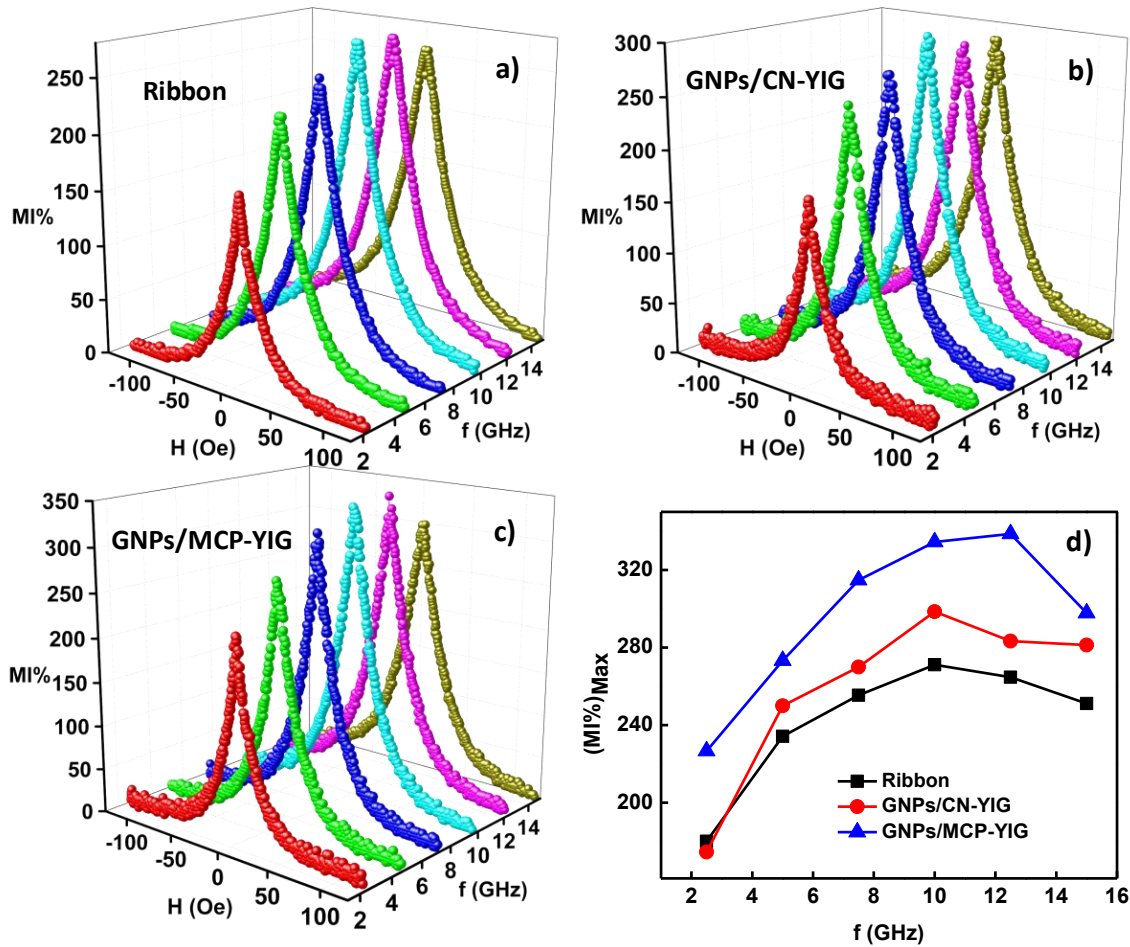

Figure 5. MI response of a) bare ribbon and ribbon drop coated by b) GNPs/MCP-YIG and c) GNPs/CN-YIG and d) the maximum of MI% for all three samples at all frequencies.

A comparison between the MI ratio of the samples at f= 10 MHz can be seen in **Figure 6**a. The amounts of MI ratio for bare ribbon, GNPs/CN-YIG coated and Gr/MCP-YIG coated ribbons were 271%, 298% and 334%, respectively. It is worth mentioning that the maximum MI% for Gr/MCP-YIG drop coated ribbon appears at 12.5 MHz. As discussed before, the reduction of MI% at high frequencies is due to presence of eddy currents that causes damping of domain wall displacements and so reduction of $\mu_t$. The impedance for the bare ribbon and the ribbon drop coated by GNPs/CN-YIG at H= 0 Oe and f=10 MHz is about 14.5 Ω. While for the ribbon drop coated with GNPs/MCP-YIG this impedance is about 13.5 Ω. In turn, the reduction of MI ratio occurs at high frequencies (lower skin depth and higher current density). Reduction of the fringe fields of the surface of the ribbon by GNPs/YIG causes significant increase of MI%. We present a tunable sensitivity by changing the strength of the magnetization of the GNPs/YIG composite. There are reports on the surface modification of magnetic ribbons by coating[27, 28, 30, 31, 45] have related this phenomenon to the closure of magnetic flux path and reduction of surface roughness. The MI field sensitivity can be defined as $\eta = d(\Delta Z/Z(\%))/d(H)$. As it can be seen in **Figure 6**b, the sensitivity is increased at the presence of composite layer on the ribbon.

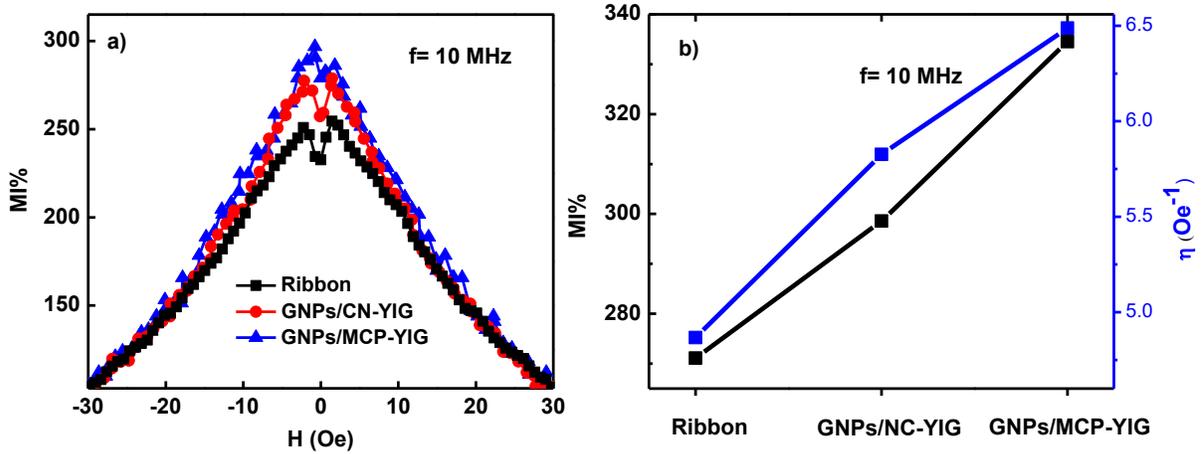

Figure 6. a) MI ratio of the samples at f= 10 MHz versus frequency. b) Maximum of MI response and field sensitivity of the samples at f= 10 MHz.

### 3.4. MI based detection of PIM in GNPs

MI is a surface sensitive effect due to the low skin depth of the ribbon. This property is the key to high functionality of MI sensors. **Figure 7** indicates the schematic of the structural conditions in MI sensor at presence of GNPs/YIG composites. Naturally, at the rough surface of the magnetic amorphous ribbons there are many fringe fields present at the surface. According to the similar amounts of the YIG concentration presented at each sample and therefore similar surface coverage, there is a similar contribution to the reduction of fringe fields from the YIG in both samples while the MI is different. Thus, we speculate that there should be another source for different MI responses between these two samples. As seen in VSM results, the proximity of GNPs to MCP-YIG yields a bigger magnetic moment in GNPs. On the other hand, the MI enhancement of ribbon for the sample coated with GNPs/MCP-YIG is higher than the ribbon coated with GNPs/CN-YIG. Therefore, according to the differences between the GNPs/CN-YIG and GNPs/MCP-YIG samples, it is deduced that PIM in GNPs is caused such a difference in the MI response of the two samples. As schematically presented in **Figure 7**, the undesired surface magnetic flux is getting diminished more in GNPs/MCP-YIG because their whole plane is being magnetized. The attenuation of the flux density on the surface of the ribbon results from a shielding effect of the ferromagnet (GNPs/YIG), which acts as a magnetic short-circuit and drives the flux lines directly towards GNPs/YIG composite. The higher the saturation magnetization the GNPs/YIG, the higher the trapped field on the top face of the ribbon and the larger the shielding effects of GNPs/YIG composite. Both of GNPs/YIG composites play this role and their MI enhanced compared to the bare ribbon, while the planar magnetized sample has more pronounced impact.

Moreover, in comparison to methods like X-ray magnetic circular dichroism (XMCD), magneto optical Kerr effect (MOKE),[46] anomalous Hall effect (AHE),[46] which provides direct proof of the magnetic proximity effect, our method presents a nearly comparative measurement tool which can be applied in a calibrated mode.

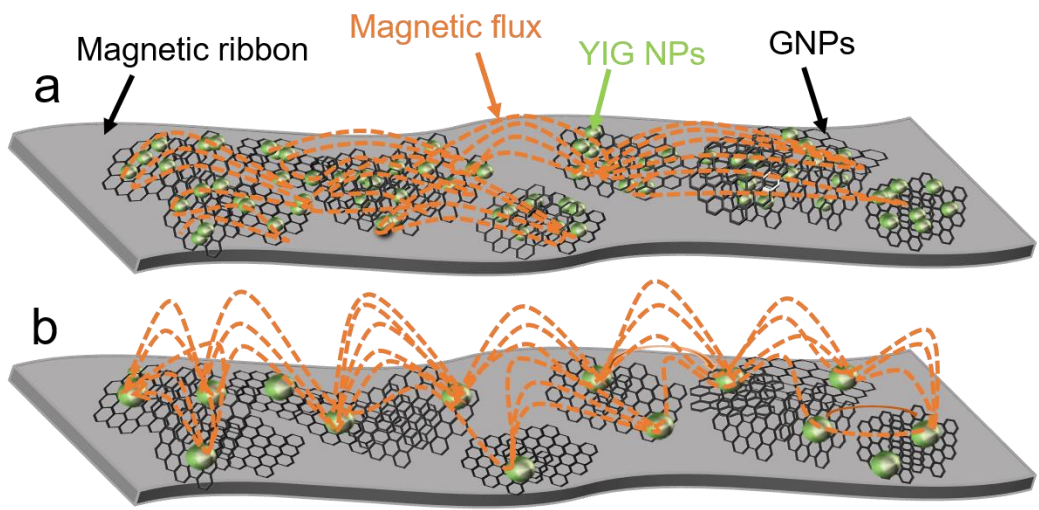

Figure 7. Schematic of the structural conditions in MI sensor and the closure of magnetic flux path when a) ribbon is coated with GNPs/MCP-YIG and b) coated with GNPs/CN-YIG. In the case of ribbon coated with GNPs/MCP-YIG, the undesired surface magnetic flux is getting diminished more, because their whole plane is being magnetized. Both of GNPs/YIG composites play this role and their MI enhanced compared to the bare ribbon, while the planar magnetized sample has more pronounced impact.

Here, according to the MI results and proximity discussions, we suggest MI sensor as a probe for measurement of PIM. According to the similarity between the VSM results of both sets of bare YIG-MNPsand large difference between their composite with GNPs, it is reasonable to derive such a conclusion. Further researches on MI can help for fully concluding this claim.

## 4. Conclusion

In summary, we have observed the PIM effect in GNPs/YIG composites. It is verified that the PIM affected by the size and clustering of YIG-MNPs, probed with microscopic observation and magnetization measurements. The higher surface area of the interface between GNPs and YIG-MNPs has resulted in the enhancement in magnetization mediated by PIM effect and also probed indirectly through the MI effect. The MI enhancement of ribbon for the sample coated with GNPs/MCP-YIG is higher than the ribbon coated with GNPs/CN-YIG. Therefore, according to the differences between the GNPs/CN-YIG and GNPs/MCP-YIG samples, it is conceived that planar magnetized GNPs has higher impact on vanishing the magnetic flux at the surface of MI element. The results of MI measurements reveal a great improvement of sensing performance and ability as a probe to detect the PIM. Our different results for PIM in GNPs/YIG composites mediated by particle size and clustering, and their effect on a proposed MI sensor can convey for further application in other magnetic systems and sensors with different mechanisms.

**Acknowledgments**

S.M. Mohseni acknowledges support from Iran Science Elites Federation (ISEF) and Iran Nanotechnology Initiative Council and Iran's National Elites Foundation (INEF).